\documentstyle[prl,preprint,aps,epsf]{revtex}
\begin{document}

\title{Randall-Sundrum Scenario at Nonzero Temperature} 
\author{
D. K. Park\raisebox{0.8ex}{1,2}\footnote[1]
{Email:dkpark@feynman.physics.lsa.umich.edu
},
Hungsoo Kim\raisebox{0.8ex}{1}\footnote[2]
{Email:hskim@hep.kyungnam.ac.kr},
Yan-Gang Miao\raisebox{0.8ex}{3,4}\footnote[3]
{Email:miao@physik.uni-kl.de},
H. J. W. M\"uller--Kirsten\raisebox{0.8ex}{3}\footnote[4]
{Email:muller1@physik.uni-kl.de}}
\address{$^1$ Department of Physics, Kyungnam University, Masan, 631-701,
	      Korea   \\
	 $^2$ Michigan Center for Theoretical Physics \\
	 Randall Laboratory, Department of Physics, University of Michigan \\
          Ann Arbor, MI 48109-1120, USA \\
	  $^3$ Department of Physics,
	  University of Kaiserslautern \\D-67653, Kaiserslautern, Germany \\
	  $^4$ Department of Physics, Xiamen University, Xiamen 361005 \\
	  People's Republic of China}

\maketitle

\maketitle
\begin{abstract}
The effect of temperature is investigated in the Randall--Sundrum
 brane--world scenario. 
It is shown that for a spacetime ansatz motivated
by similarity with AdS/CFT correspondence
several features of the model, such as its $Z_2$ symmetry, are
not maintained at nonzero temperatures.
\end{abstract}

\newpage
The Randall--Sundrum (RS) scenario was designed to solve the gauge hierarchy 
problem\cite{rs99-1}. Furthermore, the RS model involves a massless mode in the 
graviton fluctuation equation which is 
interpreted as a confined gravity
on the brane\cite{rs99-2}. In view of  its various promising features the 
RS brane--world scenario has been applied to various areas of particle
physics such as cosmology\cite{bine00,csa99-1,kanti99,csa99-2},
the  cosmological
constant problem\cite{kim00,kim01}, and black holes\cite{emp99,gid00,emp00}.
It is interesting to check whether or not
the promising features of the RS model are maintained at nonzero 
temperatures.

The most interesting feature of the RS scenario is its compatibility with
five--dimensional Einstein theory in spite of its $Z_2$ symmetry. We
first show that this remarkable feature is not maintained 
when the temperature is nonzero. 

The first step to proceed  is to examine
how the bulk spacetime is modified at finite temperature. The most
appropriate candidate can be constructed by gluing together
the two copies of the 
Schwarzschild-$AdS_5$ spacetime in a $Z_2$-symmetric manner
along a boundary which is interpreted as the
three-brane world volume:
\begin{equation}
\label{naive}
ds^2 = e^{-2k\mid y \mid}
\left[ -\left( 1 - \frac{U_T^4}{k^4} e^{4 k \mid y \mid} \right) dt^2
       +\sum^3_{i=1} dx^i dx^i \right]
       + \frac{dy^2}{1 - \frac{U_T^4}{k^4} e^{4 k \mid y \mid}}
\end{equation}
where $U_T$ is the horizon parameter and 
proportional to the external temperature
$T$ defined by $T = U_T / \pi R_{ads}$. The temperature enters
in Eq.(\ref{naive})
through the periodic identification of
$t \rightarrow t + 1 / T$ in the 
Euclidean time of
Schwarzschild-$AdS_5$ space to make the horizon at $U=U_T$
regular\cite{hawk83}.
The reason why the spacetime (\ref{naive}) is the most appropriate candidate 
for a  finite temperature RS scenario is its similarity at zero
temperature with the $AdS/CFT$ correspondence\cite{mal97,aha99}. In fact,
recently, much attention has been paid to the similarity of these somewhat
different scenarios\cite{gid00,verl99,lykk99,gub99,duff00,deg00,pere01}.
As we will show below, however, Eq.(\ref{naive}) is not a solution of the
$5d$ Einstein equation although the fine-tuning of the $5d$ cosmological
constant $\Lambda$ and the brane tension $v_b$ is appropriately adopted.

We now consider the $5d$ Einstein equation. Although the following 
analysis can also  be applied to the RS two brane
 scenario\cite{rs99-1}, we will
confine ourselves to the RS one brane scenario\cite{rs99-2} for simplicity.

The $5d$ Einstein equation in this scenario is 
\begin{equation}
\label{fieldeq}
R_{MN} - \frac{1}{2} G_{MN} R = 
- \frac{1}{4M^3} \left[\Lambda G_{MN} + v_b G_{\mu \nu}
\delta_M^{\mu} \delta_N^{\nu} \delta(y) \right]
\end{equation}
which is derived by taking the  variation of the action
\begin{equation}
\label{action}
S = \int d^4x \int dy \sqrt{-G} 
\left[-\Lambda + 2 M^3 R - v_b \delta(y) \right].
\end{equation}
In Eqs.(\ref{fieldeq}) and (\ref{action}) $M, N, \cdots$ and $\mu, \nu,
\cdots$ are $5d$ spacetime and $4d$ worldvolume indices respectively.
Motivated by Eq.(\ref{naive}) we make the {\it ansatz}
\begin{equation}
\label{ansatz}
ds^2 = e^{-2 \sigma(y)} \left[-f(y) dt^2 + dx^i dx^i\right]
+ \frac{dy^2}{f(y)}.
\end{equation}
Of course, one could choose a different ansatz for the extension
to nonzero temperatures.  But it is clearly desirable to select in the
first place one with a maximum of similarity
between the RS scenario and AdS/CFT contexts.
Then it is straightforward to show that the Einstein equation (\ref{fieldeq})
yields three independent equations:
\begin{eqnarray}
\label{relation}
6 f \sigma^{\prime 2}&-& \frac{3}{2} f^{\prime} \sigma^{\prime} = 6 k^2,
							  \\   \nonumber
f^{\prime \prime}&=& 4 f^{\prime} \sigma^{\prime},
							  \\   \nonumber
3 \sigma^{\prime \prime}&=&\frac{v_b}{4 M^3 f} \delta(y),
\end{eqnarray}
where $k^2 = -\Lambda / (24 M^3)$ and the prime denotes differentiation
with respect to $y$. If we choose $\sigma = k\mid y \mid$ and 
$f=1 - \xi e^{4 k \mid y \mid}$ as in Eq. (\ref{naive}), then the first and
third equations of (\ref{relation}) can be solved using a fine-tuning
condition $v_b = 24 M^3 k(1 - \xi)$. However, it is impossible to solve
the second equation due to a delta-function occurring in $f^{\prime \prime}$.

In fact, it may  be a formidable task to derive a solution of 
Eqs. (\ref{relation}) in a closed form. Here instead, we 
solve Eqs. (\ref{relation}) in the form of
 infinite series which is sufficient
to examine the features of the RS scenario at nonzero temperature. Motivated
by Eq. (\ref{naive}) again, we introduce a small dimensionless 
parameter $\xi$ as follows:
\begin{eqnarray}
\label{intro}
f&=&1 + \xi f_1(y) + \xi^2 f_2(y) + \cdots,   \\   \nonumber
\sigma&=&k \mid y \mid + \xi \sigma_1(y) + \xi^2 \sigma_2(y) + \cdots.
\end{eqnarray}
The expansion parameter $\xi$ is related to the external temperature
of our universe, and the explicit relation will be derived 
later(see Eq.(\ref{temdef})).

Inserting Eq.(\ref{intro}) into the first two equations of (\ref{relation})
one can solve these for $f_i$ and $\sigma_i$ step by step.
 First we note the equation
\begin{equation}
\label{note}
f_1^{\prime \prime} = 4 k \epsilon(y) f_1^{\prime}
\end{equation}
which originates  from $O(\xi)$ contributions in the second 
of equations (\ref{relation}).
Eq.(\ref{note}) implies $f_1^{\prime} \propto e^{4 k \mid y \mid}$
which means that
$f_1$ is $Z_2$--antisymmetric. If one chooses $f_1 = 0$ 
in an attempt to maintain the $Z_2$ symmetry of the original zero temperature
RS scenario,
 the higher order equations imply that all $f_i$ and
$\sigma_i$ are zero. Thus we return to the original zero-temperature
RS scenario. 
The other possibility of choosing  $\sigma =k|y|$ implies from the
first of eqs. (\ref{relation}) that
$f^{\prime}_i(y)\delta (y)=0$ and so has a similar effect.
We therefore
 abandon the $Z_2$ symmetry at finite temperature. It is relatively
straightforward to derive now $f_i$ and $\sigma_i$ perturbatively. Here, we 
present the first few solutions explicitly:
\begin{eqnarray}
\label{solu}
f_1&=& -\frac{1}{4} \epsilon(y) (e^{4 k \mid y \mid} - 1),
\hspace{2.0cm} 
f_2 = \frac{1}{8} (k\mid y \mid - \frac{1}{4}) e^{4 k \mid y \mid},
					    \\    \nonumber
f_3&=&-\frac{1}{32} \epsilon(y)
\left[(k^2 y^2 + \frac{1}{4} k \mid y \mid) e^{4 k \mid y \mid}
      - \frac{1}{16} (e^{4 k \mid y \mid} - 1)  \right],
                                             \\    \nonumber
f_4&=& \frac{k^2 y^2}{384} (3 + 2 k \mid y \mid) e^{4 k \mid y \mid},
					     \\     \nonumber
\sigma_1&=&-\frac{k}{8} y   
\hspace{2.0cm}
\sigma_2 = \frac{3k}{128} \mid y \mid,
					      \\     \nonumber
\sigma_3&=&-\frac{k}{256} y,
\hspace{2.0cm}
\sigma_4 = \frac{23 k}{32768} \mid y \mid.
\end{eqnarray}
Then the third of Eqs.(\ref{relation}) yields an additional 
fine-tuning condition
\begin{equation}
\label{fine}
v_b = 24 M^3 k \left[ 1 - \frac{\xi^2}{128} - \frac{\xi^4}{32768} +
		      \cdots \right].
\end{equation}

A further interesting feature of the RS
scenario is that its spacetime consists
of two copies of $AdS_5$ space. Due to $Z_2$--symmetry breaking, however, 
the two spaces attached to the boundary are not necessarily identical
at finite temperature. In order to show this
explicitly, we define a 
variable 
\begin{equation}
\label{def1}
z \equiv \frac{1}{k} e^{\sigma(y)}
= \frac{1}{k} e^{k\mid y \mid}
\left[1 - \frac{\xi}{8} ky + \frac{\xi^2}{128} (3k\mid y \mid + k^2 y^2) +
      \cdots   \right].
\end{equation}
In terms of $z$ the spacetime (\ref{ansatz}) becomes
\begin{equation}
\label{stime1}
ds^2 = \frac{1}{k^2 z^2}
\left[-f dt^2 + dx^i dx^i + \frac{k^2}{f \sigma^{\prime 2}} dz^2 \right].
\end{equation}
Inverting Eq.(\ref{def1}) one can show straightforwardly that
\begin{eqnarray}
\label{exp1}
f&=&(1 + \frac{\xi}{4}) \left[1 - (\frac{\xi}{4} - \frac{\xi^2}{32} + \cdots)
			      (kz)^4 \right],
						     \\   \nonumber
f\sigma^{\prime 2}&=& k^2 \left[1 + O(\xi^3) \right]
\left[1 - (\frac{\xi}{4} - \frac{\xi^2}{32} + \cdots)
			      (kz)^4 \right]
\end{eqnarray}
for the case of positive $y$. Expansion in the case of
negative $y$ results in the same 
Eq.(\ref{exp1}) with $\xi$ changed into $-\xi$. If one defines a 
new time variable $t_{\pm} \equiv \sqrt{1 \pm \frac{\xi}{4}} t$ for
$y > 0$ and $y < 0$ respectively, the final form of spacetime becomes
\begin{equation}
\label{finalstime}
ds^2 = \frac{1}{k^2 z^2}
\left[-(1 - U_T^{(\pm) 4} z^4) dt_{\pm}^2 + dx^i dx^i + 
      \frac{dz^2}{1 - U_T^{(\pm) 4} z^4}  \right]
\end{equation}
where
\begin{equation}
\label{temdef}
U_T^{(\pm) 4} = \pm k^4 (\frac{\xi}{4} \mp \frac{\xi^2}{32} + \cdots).
\end{equation}
If we choose $\xi(y>0) = -\xi(y<0)$, Eq.(\ref{finalstime}) represents
two copies of Schwarzschild--$AdS_5$ space. In this case, however, 
Eq.(\ref{solu}) is no longer a solution of Einstein's
 equation. If, on the other hand,
we choose $\xi$ to be constant, {\it say} $\xi > 0$, the space in the $y > 0$
region is a usual Schwarzschild--$AdS_5$ space,
 but the space in the region $y < 0$ is
not, due to $U_T^{(-) 4} < 0$. Although this is also obtained from
a decoupling limit of the black three-brane solution\cite{horo91}
by interchanging
 the inner and outer horizons, it is unclear why the
RS scenario requires this apparant asymmetry. Eq.(\ref{temdef}) allows one to
express $\xi$ in terms of the external temperature. 

Finally, we examine  features of the RS scenario at the level of the 
four-dimensional effective action. In the original RS scenario the 
fine--tuning condition $\Lambda = -24 M^3 k^2$ and $v_b = 24 M^3 k$ enables
one to derive a four--dimensional effective action with a vanishing cosmological
constant from a wide range of five--dimensional metrics. One can show this
explicitly by computing a $4d$ effective action from a $5d$ line
element
\begin{equation}
\label{wrstime}
ds^2 = e^{-2k\mid y \mid} g_{\mu \nu}(x) dx^{\mu} dx^{\nu} + dy^2.
\end{equation}
Of course, Eq.(\ref{wrstime}) agrees with the RS solution when 
$g_{\mu \nu} = \eta_{\mu \nu}$. 

At a finite temperature, however, this kind of feature seems to be severely 
restricted. In order to see this more explicitly 
we introduce a line
element
\begin{equation}
\label{line1}
ds^2 = e^{-2 \sigma(y)} g_{\mu \nu}(x, y) dx^{\mu} dx^{\nu} + 
\frac{dy^2}{f(y)}
\end{equation}
where
\begin{equation}
\label{metric1}
g_{\mu \nu}(x, y) = \bar{g}_{\mu \nu}(x) + (1 - f(y)) \bar{g}_{\mu}^0
						      \bar{g}_{\nu}^0.
\end{equation}
Here, $\bar{g}_{\mu \nu}(x)$ represents a physical graviton in the effective 
theory. The curvature scalar $R$ computed from the metric (\ref{line1}) can be 
shown to be
\begin{equation}
\label{curvature}
R = e^{2 \sigma} \bar{R} + \Delta R,
\end{equation}
where $\bar{R}$ is the $4d$ curvature scalar derived from $\bar{g}_{\mu \nu}$ 
and
\begin{equation}
\label{remain}
\Delta R = \frac{2 v_b}{3 M^3} \delta(y) - 20 k^2
- (1 + \bar{g}^{00}) 
\left[ \frac{f^{\prime}\sigma^{\prime}}{1 + (1 - f) \bar{g}^{00}}
    - \frac{\bar{g}^{00} f^{\prime 2}}{2 [1 + (1 - f) \bar{g}^{00}]^2}
							      \right].
\end{equation}
One should note the following.
 In Eq.(\ref{curvature})
we have dropped several terms which are irrelevant for the following
discussion. In fact, these terms become zero for the constant metric
of the worldvolume which is what we are interested in.
 Also we have changed $\Delta R$ into a
more convenient form for the following discussion using Eq.(\ref{relation}).

Using Eq.(\ref{curvature}) and 
\begin{equation}
\label{det}
\sqrt{-G} = \sqrt{-\bar{g}_4} e^{-4 \sigma} 
\sqrt{\frac{1 + (1 - f) \bar{\mu}}{f}}
\end{equation}
where $\bar{g}_4 = det \bar{g}_{\mu \nu}$, $\bar{g}_3 = det \bar{g}_{ij}
(i,j=1,2,3)$, and $\bar{\mu} \equiv \bar{g}_3 / \bar{g}_4$, one can 
calculate a four-dimensional effective action whose form is 
\begin{equation}
\label{effaction}
S_{eff} = \int d^4 x \sqrt{-\bar{g}_4} [2 M_{pl}^2 \bar{R} - \Lambda_4]
\end{equation}
where $M_{pl}^2 = M^3 / k$ and $\Lambda_4$ is four-dimensional cosmological
constant. 

For the Ricci-flat case $\Lambda_4$ assumes the form
\begin{equation}
\label{cosmoc}
\Lambda_4 = \int dy e^{-4 \sigma} 
\sqrt{\frac{1 + (1 - f) \bar{\mu}}{f}}
\left[ \Lambda + v_b \delta(y) - 2 M^3 \Delta R_1 \right].
\end{equation}

We now consider a flat space case by taking
\begin{eqnarray}
\label{flatlimit}
\bar{\mu}&=&-1,        \\   \nonumber
\Delta R_1&=&\frac{2 v_b}{3 M^3} \delta(y) - 20 k^2.
\end{eqnarray}
Then it is easy to show that $\Lambda_4$ in flat space becomes
\begin{equation}
\label{flatcos}
\Lambda_4^{flat} = 16 M^3 k^2 \int dy e^{-4 \sigma} - \frac{v_b}{3}.
\end{equation}
Although not obvious at this stage, one can show as follows
that the right--hand side of
Eq.(\ref{flatcos}) is zero. After expanding $e^{-4 \sigma}$ in 
terms of $\xi$, integration with respect to $y$ yields
\begin{equation}
\label{exp2}
\int dy e^{-4 \sigma} = \frac{1}{2k}
\left[1 - \frac{\xi^2}{128} - \frac{\xi^4}{32768} + \cdots \right]
\end{equation}
which indicates $\Lambda_4^{flat} = 0$ via the fine--tuning 
condition (\ref{fine}).

However, the vanishing of the cosmological constant is not maintained when the 
time-time component of $\bar{g}_{\mu \nu}$ deviates from Lorentz
signature $-1$. To show this we choose 
$\bar{g}^{\mu \nu} = \eta^{\mu \nu} + \beta \eta_0^{\mu} \eta_0^{\nu}$. Then,
$\bar{\mu}$ and $\Delta R$ become
\begin{eqnarray}
\label{devi1}
\bar{\mu}&=& 1 - \beta,     \\   \nonumber
\Delta R&=& \frac{2 v_b}{3M^3} \delta(y) - 20k^2 -
\beta \left[\frac{f^{\prime}\sigma{\prime}}{f + \beta (1-f)} + 
\frac{(1-\beta) f^{\prime 2}}{2 [f + \beta (1-f)]^2} \right].
\end{eqnarray}
Expanding the integrand in Eq.(\ref{cosmoc}) in terms of $\xi$ again, 
it is straightforward to show that
\begin{equation}
\label{devi2}
\Lambda_4^{devi} = \Lambda^{(0)} + \Lambda^{(2)} \xi^2 + \cdots
\end{equation}
where $\Lambda^{(0)} = 0$ and 
\begin{equation}
\label{devi3}
\Lambda^{(2)} = M^3 k \beta
\left[\frac{3 \beta - 5}{8} - \frac{1}{2} (kL) - \frac{3}{4} (kL)^2
      + \frac{3 - 2 \beta}{4} e^{2kL} \right]
\end{equation}
where $L$ is the length of the fifth dimension. Although $\Lambda^{(2)}$ is
infinite due to the appearance of $L$, it may give a finite, but 
$r_c$-dependent cosmological constant if the same procedure is applied to
the RS two brane scenario\cite{rs99-1}.  

In summary,  we have examined the RS scenario when the external temperature 
is nonzero by using a spacetime ansatz motivated by 
analogy with that of AdS/CFT correspondence. 
We have shown that in such a case 
many interesting features of the original 
RS model are not 
maintained at nonzero temperatures. It is interesting to check whether
or not the finite temperature solution (\ref{intro}) and (\ref{solu}) also
has a massless bound state at the fluctuation level. The more important
point, however, is to reformulate the finite temperature RS scenario
from the string/$M$ theory background. Then, it may be possible to 
find an answer why many interesting features of the RS model disappear at 
nonzero temperature.

\vspace{1cm}

{\bf Acknowledgement}: DKP acknowledges support  
from the Basic Research Program of the Korea
Science and Engineering Foundation (Grant No. 2001-1-11200-001-2). YGM
acknowledges supports by an Alexander von Humboldt fellowship, by the
National Natural Science Foundation of China(Grant No. 19705007), and
by the Ministry of Education of China under the special project for scholars 
returned from abroad.

\end{document}